\documentclass[11pt]{article}
\usepackage{epsfig,multicol}
\usepackage[english]{babel}
\usepackage{graphicx}
\usepackage{rotating}
\usepackage{amsmath}
\usepackage{amssymb}
\usepackage{flafter}

\def\sss{\scriptscriptstyle}
\def\brmu{{\rm BR}(\mu\rightarrow e\gamma)}
\def\ie{{\em i.e.}}

\begin{document}

\begin{titlepage}

\begin{flushright}
SISSA 65/2003/EP 
\end{flushright}

\begin{center}

\vspace*{3cm}
{\LARGE Non-Universal Gaugino Masses and the Fate of $\mu\rightarrow e\gamma$}\\
\vspace*{1.5cm}
{\bf\large Stefano Profumo and Carlos E. Yaguna}\\[0.7cm]
        {\em Scuola Internazionale Superiore di Studi Avanzati,
	Via Beirut 2-4, I-34014 Trieste, Italy \\
and Istituto Nazionale di Fisica Nucleare, Sezione di Trieste, 
I-34014 Trieste, Italy\\[0.3cm] 
	E-mail: {\tt profumo@sissa.it, yaguna@sissa.it} 
} 

\vspace*{0.8cm}


\begin{abstract}We study charged lepton flavor violating processes in the context of supersymmetric models with non-universal gaugino masses. We consider the generic case of free gaugino masses, as well as specific models based on gaugino mediation and $SU(5)$ GUTs. Under the assumption that the main source of lepton flavor violation resides in the off-diagonal elements of the left handed slepton mass matrix, we show that $\brmu$ strongly depends on gaugino non-universality. When $M_1$ and $M_2$ have opposite signs, cancellations occurring between the neutralino and the chargino contributions can suppress, even completely, the branching ratio of charged lepton flavor violating decays. Moreover, we find that for \mbox{$M_3> M_2,M_1$}, $\brmu$ is an increasing function of $M_3$.
\end{abstract}  
\end{center}

\end{titlepage}



\section{Introduction}
Universal gaugino masses are not a consequence of the supergravity framework but rather an additional assumption. In supergravity models, supersymmetry is broken in a hidden sector which consists of fields which couple to the MSSM fields only {\em gravitationally}. In the minimal scenario (mSUGRA), it is assumed that the vacuum expectation value of the gauge kinetic function does not break the unifying gauge symmetry. This simplification leads to universal gaugino masses. In mSUGRA it is also assumed that all  scalars acquire a common mass, and that the trilinear couplings are universal. Such simplifying assumptions, however, are not necessarily valid in specific scenarios. Indeed, in the literature there is plenty of models which do not fulfill such requirements \cite{Baer:2000gf}. It is then worth to consider the phenomenological implications of non minimal scenarios, and in particular of gaugino non-universality \cite{ref:gaugnonuniv}.

So far, the discussion of models with non-universal gaugino masses has focused on the implications for SUSY particles detection in accelerator experiments \cite{Krasnikov:2001pd} and dark matter \cite{ref:dmgnu}. One of the features of these models is that the cosmological upper bound on the relic density \cite{Spergel:2003cb} can be easily fulfilled. In fact, the lightest supersymmetric particle may be a wino- or higgsino-dominated rather than a bino-dominated neutralino. In both cases, annihilations into $W^+W^-$ and $Z^0Z^0$, as well as coannihilations with the lightest chargino and (for the higgsino) with the next-to-lightest neutralino can suppress the relic density below the currently estimated dark matter content of the Universe \cite{Bertin:2002sq,Chattopadhyay:2003yk}.  Additionally, such  neutralinos, due to their large couplings to the $W$ and $Z$ bosons, typically yield large direct and indirect detection rates, and therefore constitute appealing candidates for WIMP search experiments \cite{Bertin:2002sq,Chattopadhyay:2003yk,Birkedal-Hansen:2002am}. Lepton flavor violating (LFV) decays, on the other hand, have never been studied in this context.

The process $\mu\rightarrow e\gamma$ is an ideal candidate to look for new physics \cite{LFV, hisano}. Recent experiments on neutrino oscillations have shown that flavor is not conserved in the lepton sector, implying that lepton flavor violating decays such as $\mu\rightarrow e\gamma$, $\tau\rightarrow \mu\gamma$ and $\mu\rightarrow 3e$ are in principle allowed. Since in standard model-like theories these processes are extremely suppressed, their detection would be an unmistakable signal of physics beyond the Standard Model. In a supersymmetric framework, the precise rate of LFV decays depends on the mechanism that communicates the violation of lepton flavor to the soft breaking Lagrangian. To maintain our discussion as general as possible, instead of studying one of the different scenarios considered in the literature, we will conveniently parametrize the terms assumed to be responsible for LFV at low energies.
 
In this article we study the effects that non universal gaugino masses can have in the decay $\mu\rightarrow e\gamma$. In the next section we explain our assumptions and introduce some notation. Then, we briefly review the predictions for $\brmu$ in models with universal gaugino masses, and afterwards we proceed to consider general models with non universality, which we study in some detail. Finally, we discuss models based on the unifying group $SU(5)$, which predict a fixed pattern of non-universality, and models with gaugino mediated SUSY breaking.

\section{Generalities}
A generic model with non universal gaugino masses contains, in addition to the mSUGRA parameters $m_0,a_0,\tan\beta$ and $\mathrm{sign}(\mu)$, three gaugino masses $M_1,M_2,M_3$ corresponding, respectively, to the $U(1)$, $SU(2)$ and $SU(3)$ gauge groups. The effect of non universality can be parametrized, at the high energy GUT scale $M_{\sss\rm GUT}\simeq2\ 10^{16}$ GeV, by the dimensionless ratios $r_2=M_2/M_1$ and $r_3=M_3/M_1$. In the following, we will study the dependence of $BR(\mu\rightarrow e\gamma)$ on $r_2$ and $r_3$ for different values of the remaining parameters. 

At low energies, the only possible sources of lepton flavor violation in the MSSM soft breaking Lagrangian are non-vanishing off-diagonal elements in $m_{\sss\widetilde L}^2$, $m_{\widetilde e}^2$ and $A_e$. We will assume that the dominant source of LF violation  resides in the left-handed slepton mass matrix $m_{\sss\widetilde L}^2$, and parametrize the off-diagonal element $(m_{\sss\widetilde L}^2)_{21}$ in terms of the diagonal entry as\begin{equation} 
(m_{\sss\widetilde L}^2)_{21}=\delta_{\sss\rm OD}(m_{\sss\widetilde L}^2)_{11}=\delta_{\sss\rm OD}(m_{\sss\widetilde L}^2)_{22}, 
\end{equation}
$\delta_{\sss\rm OD}$ being a free parameter. The off-diagonal elements $(m_{\sss\widetilde L}^2)_{13}$ and $(m_{\sss\widetilde L}^2)_{23}$ play a subdominant role in the process $\mu\rightarrow e\gamma$, since they only contribute through double mass insertions, and will therefore be disregarded in what follows. The matrices  $m_{\widetilde e}^2$ and $A_e$, on the other hand, are taken to be diagonal. The outlined scenario applies, for instance, in see-saw inspired models of LFV \cite{Hisano:1995nq,LFV}. We remark that, though our results refer to the decay $\mu\rightarrow e\gamma$, with obvious substitutions, they  also hold for the decays $\tau\rightarrow\mu\gamma$ and  $\tau\rightarrow e\gamma$.

The amplitude for the process $\mu\rightarrow e\gamma$ is written
\begin{equation}
T=e\epsilon^{\alpha *}\bar u_e(p-q)\left[m_{\mu}i\sigma_{\alpha\beta}q^\beta(A^L P_L+A^R P_R)\right]u_\mu(p)
\end{equation}
where $q$ is the momentum of the photon, $e$ is the electric charge, $\epsilon^*$ the photon polarization vector, $u_e$ and $u_\mu$ are the wave functions for the electron and the muon, and $p$ is the momentum of the muon. Each coefficient in the above equation can be written as a sum of two terms
\begin{equation}
A^{L,R}=A^{(n)L,R}+A^{(c)L,R}
\end{equation}
where $A^{(n)L,R}$ and $A^{(c)L,R}$ stand, respectively, for the contribution from the neutralino-slepton and the chargino-sneutrino loops. Explicit  expressions for these contributions can be found in \cite{hisano}. The branching ratio $\mu\rightarrow e\gamma$ is given by
\begin{equation}
\brmu=\frac{48\pi^3\alpha}{G_F^2}(|A^L|^2+|A^R|^2)\,.
\end{equation}
Owing to the hierarchy of masses in the lepton sector ($m_\mu\gg m_e$) the right-handed term always dominates, $|A^R|^2\gg|A^L|^2$. Hereafter, we will refer to $A^{(c)R}$ and $A^{(n)R}$ as the {\em chargino} and {\em neutralino} contributions respectively.

\section{The Universal Case}\label{sec:universal}

\begin{figure*}[tb]
\begin{center}
\begin{tabular}{ccc}
\includegraphics[scale=0.5]{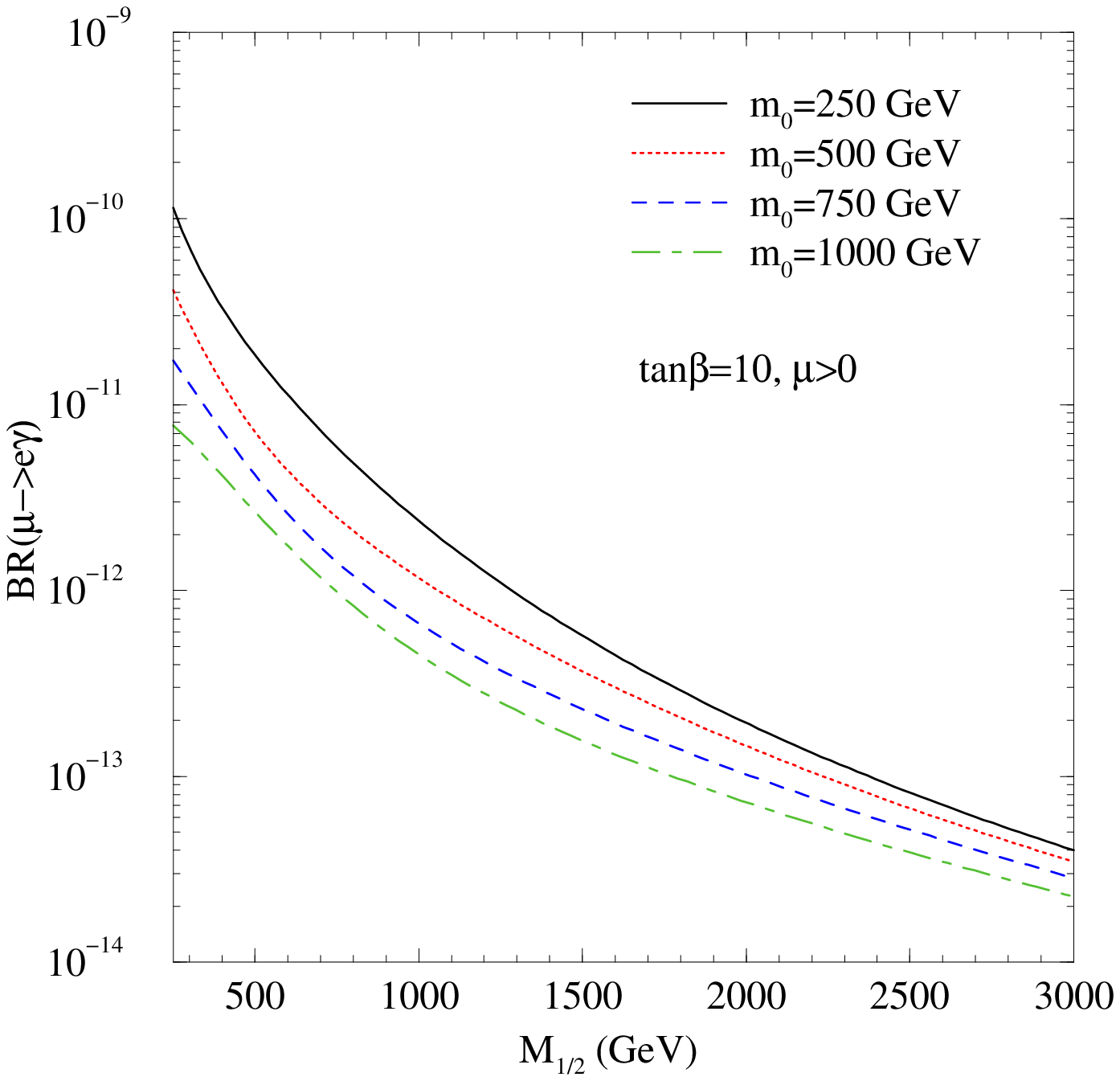}& \qquad &
\includegraphics[scale=0.5]{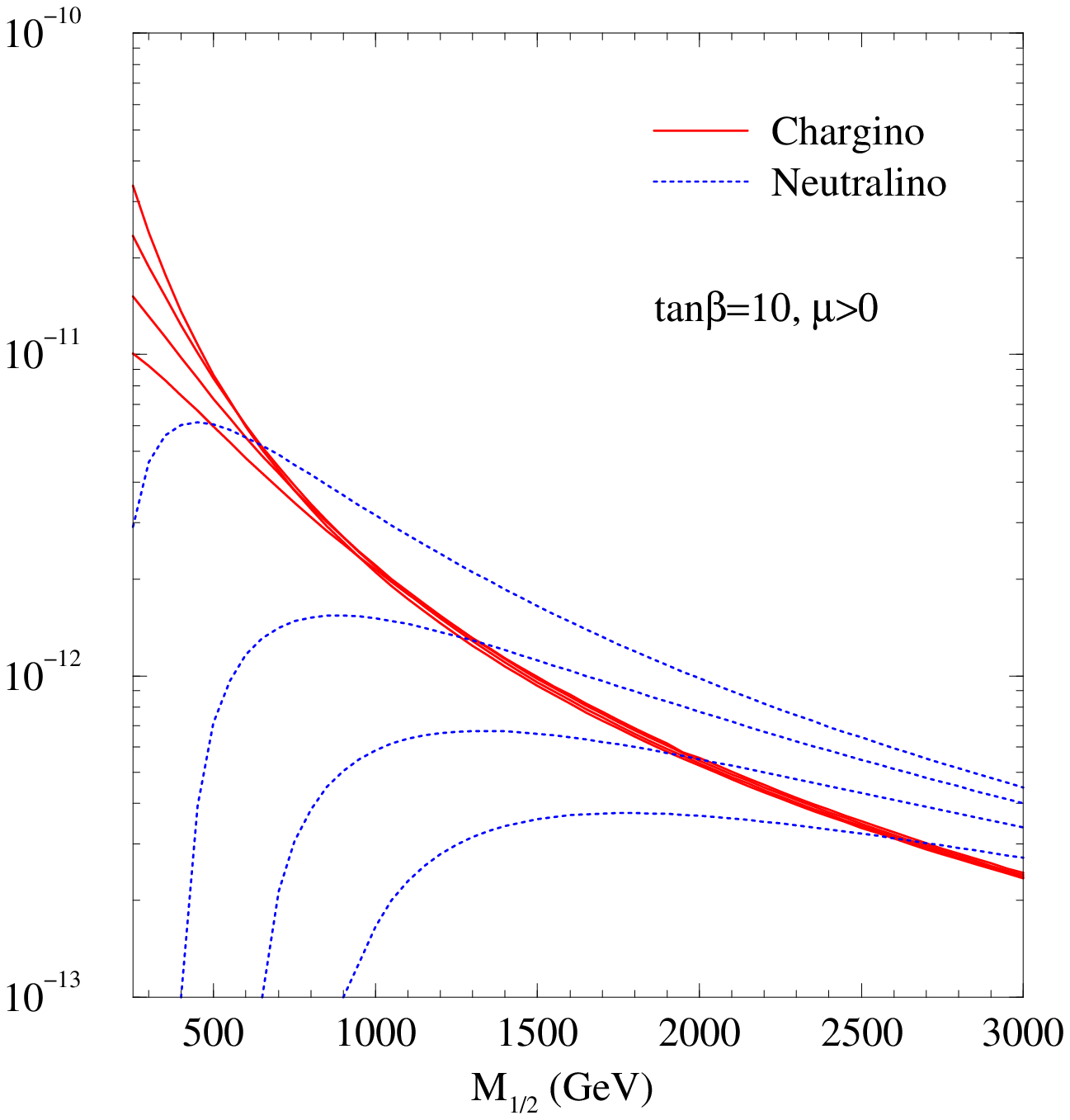}\\
\end{tabular}
\end{center}
\caption{\label{fig:UNIVERSAL}: The $\brmu$ ({\em left} ) and the chargino-sneutrino ({\em right}, solid red lines) and neutralino-slepton  ({\em right}, dashed blue lines) contributions as functions of the common gaugino mass $M_{1/2}$ for $m_0=250,\ 500,\ 750$ and 1000 GeV. $A_0=0$, $\tan\beta=10$ and $\mu>0$.}
\end{figure*}

In the framework of the constrained MSSM, the occurrence of off diagonal entries in the slepton mass matrix $m^2_{\sss\widetilde L}$ as introduced in the current setup, gives rise to charged lepton flavor violating decays of the type $BR(l_i\rightarrow l_j\gamma)$. The corresponding rates depend in general on the supersymmetric parameters of the theory, which, we recall, are given on the one side by the GUT-scale inputs $m_0$, $M_{1/2}$, $A_0$ and on the other side by $\tan\beta$, the sign of $\mu$ and the relative size of the relevant off-diagonal entry of ($m_{\sss\widetilde L}^2$), which we denote as $\delta_{\rm\sss OD}$. In this setting, $BR(l_i\rightarrow l_j\gamma)$ is, to a good accuracy, {\em proportional} to $\tan\beta$, while the dependence on the sign of $\mu$ and on $A_0$ is rather weak\footnote{See e.g. the discussion in \cite{Petcov:2003zb}.}. As regards the mass scales of the theory, $M_{1/2}$ and $m_0$, it is clear that {\em larger masses} will in general {\em suppress} lepton flavor violating decays, since particles circulating into the loops are heavier. In particular, being the masses of the charginos and neutralinos mostly insensitive to $m_0$, the strongest dependence will be that on $M_{1/2}$. 

We plot in the left panel of fig.~\ref{fig:UNIVERSAL} the results we get for $\brmu$ as a function of $M_{1/2}$, in the CMSSM scenario for $\delta_{\sss\rm OD}=10^{-3}$, at $\tan\beta=10$, $\mu>0$ at various values of $m_0$. Henceforth, where not otherwise specified, $A_0$ is set to zero and $\delta_{\sss\rm OD}=10^{-3}$, for definiteness. We verified that, within the present setup, the dependence of LFV branching ratios on $A_0$ is always negligible.

In the right panel of fig.~\ref{fig:UNIVERSAL} we split the contributions arising from the chargino and the neutralino loops. We observe that, as a rule, the chargino contribution is larger than that of the neutralino for sufficiently low values of $M_{1/2}$, while, {\em vice-versa} the neutralino may dominate at large $M_{1/2}$. Moreover, the chargino contribution is almost independent of $m_0$, on which instead the neutralino contribution rather critically depends. This behavior stems from the {\em steepness} of the different functions entering $A^{(c)}$ and $A^{(n)}$, whose arguments are the ratios $(m^2_{\sss\tilde{\chi}^-_A}/m^2_{\tilde\nu})$ and $(m^2_{\sss\tilde{\chi}^0_A}/m^2_{\tilde l})$, in the region where they are evaluated. 

Finally, we stress that the two contributions have the {\em same sign}, possibly excluding a narrow region at very low $M_{1/2}$, where the chargino contribution is however always by far dominant. Flipping the sign of $\mu$ would change the sign of both contributions simultaneously. Therefore, in the universal case {\em no cancellations} among the chargino and the neutralino contribution can occur.

\section{Non-universal Wino Mass Term $M_2$}

\begin{figure*}[t]
\begin{center}
\includegraphics[scale=0.5]{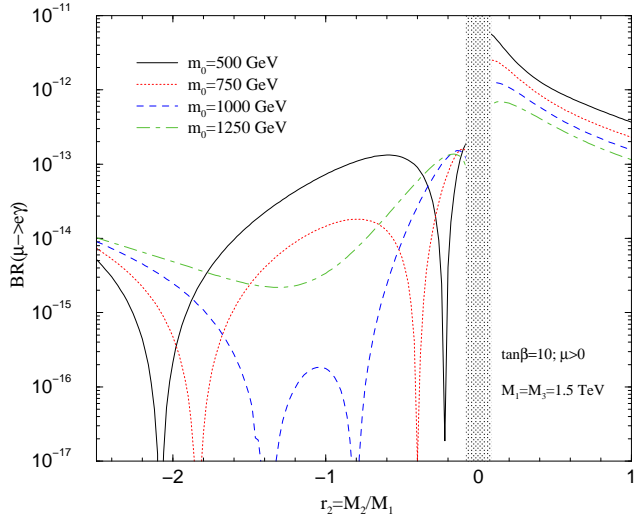}
\end{center}
\caption{\label{fig:BRr2} The $\brmu$ as a function of the ratio $r_2\equiv M_2/M_1$ for $M_1=M_3=1.5\ {\rm TeV}$ at $\tan\beta=10$, $\mu>0$, for $m_0=500,\ 750,\ 1000$ and 1250 GeV.}
\end{figure*}

In this section we focus on the effects, on LFV processes, of taking, at the unification scale $M_{\sss\rm GUT}$, the gaugino mass pattern $M_1=M_3$, $M_2=r_2 M_1$, and varying $r_2$. The effect of a non-universality in the high energy input value of the wino mass term mainly translates, at the low energy scale, into a proportional variation of $M_2$, and thus of the wino content of both the charginos and the neutralinos \cite{Birkedal-Hansen:2001is}. We remark that, through the electroweak symmetry breaking conditions, however, also the value of $\mu$ is slightly sensitive to $r_2$.

We start our analysis studying, in fig.~\ref{fig:BRr2}, the branching ratio $\brmu$ as a function of $r_2$, at fixed $M_1=M_3=1500\ {\rm GeV}$ and $\tan\beta=10$. The sign of $\mu$ is chosen to be positive, and the various lines correspond to different choices of $m_0$, ranging from 500 GeV up to 1250 GeV. The shaded region around $r_2=0$ signals a chargino mass which is not compatible with the experimental lower bound $m_{\tilde\chi^-}\gtrsim 100\ {\rm GeV}$ from LEP \cite{Hagiwara:fs}. The extremum right hand-side of the plot represents the particular point at $r_2=1$ corresponding to gaugino mass universality. We notice that no particular effect, a part the obvious suppression of $\brmu$ due to a heavier chargino and neutralino spectrum for increasing $|r_2|$, is present in the region where the sign of $M_2$ is the same as the sign of $M_1=M_3$ (\ie\ $r_2>0$). Contrary, for negative $r_2$, cancellations may occur between the chargino and the neutralino contributions. When the two contributions are exactly equal and opposite, $\brmu$ drops to zero, as signaled by the sharp dips appearing in the figure. In the region between two dips the dominant contribution is that of the neutralino, while for $r_2\rightarrow0$ and $r_2\rightarrow\infty$ the chargino contribution always dominates. We emphasize that such cancellations {\em would not be spoiled} by the presence of a subdominant contributing amplitude, coming e.g. from $A_e$, $m_{\widetilde e}$ or double mass insertions. The only net effect would be a small shift in the precise location of the dips, which would be nonetheless present.

\begin{figure*}
\begin{center}
\includegraphics[scale=0.5]{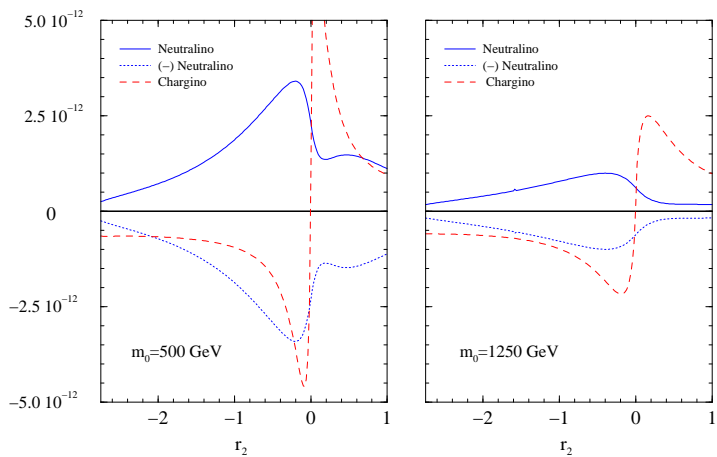}
\end{center}
\caption{The separate chargino and neutralino contributions to $\brmu$ in the two cases $m_0=500$ and 1250 GeV shown in fig.~2.}\label{fig:r2CONTRIB}
\end{figure*}
Fig.~\ref{fig:r2CONTRIB} shows the separate contributions from the chargino and the neutralino for the two extreme cases $m_0=500$ and 1250 GeV. For the sake of clarity, we also plot the neutralino contribution with opposite sign, in order to highlight the points where the cancellations take place for $m_0=500$ GeV. The reason of the wide dip in the $m_0=1250$ GeV case is also apparent from the right panel.

The location of the dips depends rather sensitively on the value of $m_0$, and in some cases it may take place in the cosmologically interesting region where the neutralino is wino-like. In some cases (as for $m_0$=1250 GeV in the figure) the exact cancellation never takes place, though it is evident that the chargino and neutralino loops interfer destructively, generating a minimum at some value of $r_2$, contrary to what one would expect from the overall increase of the mass spectrum.

\begin{figure*}
\begin{center}
\includegraphics[scale=0.55]{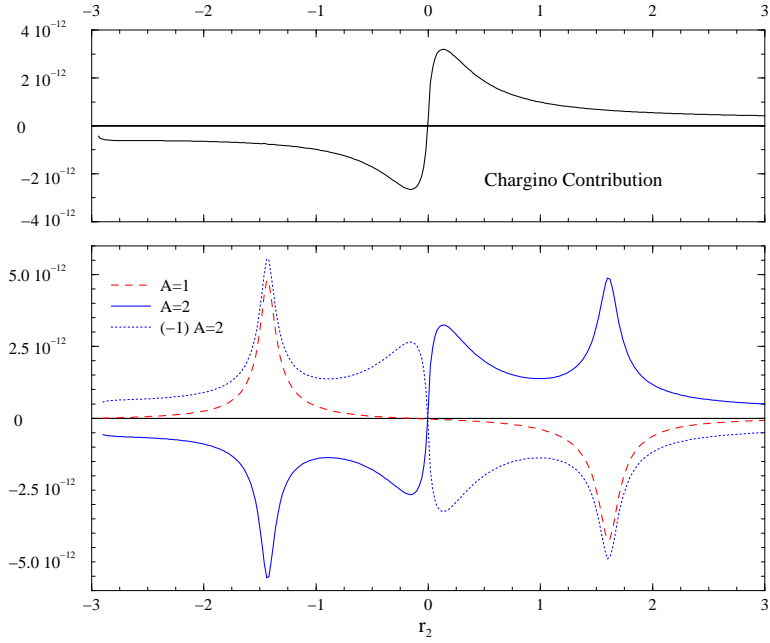}\\
\caption{({\em Upper panel} ): The chargino contribution as a function of $r_2$ at $M_1=M_3=1.5\ {\rm TeV}$, $\tan\beta=10$, $\mu>0$ and $m_0=1000$. ({\em Lower panel} ): the contributions from the light ($A=1$) and heavy ($A=2$) chargino loops at the same supersymmetric parameters as above. Also included is the $A=2$ contribution with changed sign.}\label{fig:CHCONTRIB}
\end{center}
\end{figure*}

The chargino contribution depends on the orthogonal matrices $O_L$ and $O_R$ which diagonalize the chargino mass matrix $M_C$. In particular, we define $O_R$ as the orthogonal matrix which diagonalizes $M_C M_C^T$, and $O_L$ as that which diagonalizes the combination $M_C^T M_C$. We call {\em chargino masses} the two diagonal elements $M_{\tilde\chi^-_A}$ of the matrix $O_R M_C O_L^T$, which can therefore also take {\em negative values} \cite{Haber:1984rc}. Multiplication of $O_L$ by the Pauli matrix $\sigma_3$ allows to recover the positive values for the masses. We recall that the diagonal entries of $M_C$ are $M_2$ and $\mu$, therefore a large mixing, and sizable off-diagonal entries in $O_{L,R}$, appear when $M_2=\pm\mu$. Since the largest element in the chargino amplitude is given by 
\begin{equation}\label{eq:CHRGCONTRIBUTIONS} 
A^c\simeq -{\rm const}\cdot \sum_{A,X}\ (O_R)_{\sss A1}(O_L)_{\sss A2}\frac{M_{\tilde\chi^-_A}}{m_\mu}U^\nu_{X1}U^\nu_{X2} C(x_{\sss AX}),\quad x_{\sss AX}\equiv M^2_{\tilde\chi^-_A}/m^2_{\tilde\nu_X},
\end{equation}
we expect an enhancement in $A^{(c)}$ when  $M_2^2=\mu^2$, being the amplitude always proportional to an off diagonal element of $O_{L,R}$. Though a peak indeed takes place at \mbox{$M_2=\pm\mu$} (around $r_2\simeq\pm1.4$ in fig.~\ref{fig:CHCONTRIB}) in the separate contributions from the lighter chargino ($A=1$) and from the heavier chargino ($A=2$), the overall result, shown in the upper panel, is instead completely smooth\footnote{Enhancements at $M_2=\pm\mu$ have instead been reported in other related frameworks, such as in supersymmetric electroweak baryogenesis \cite{Brhlik:1999qr}}. In order to understand why the peaks are smoothed out in the total chargino contribution, a closer look at the amplitude (\ref{eq:CHRGCONTRIBUTIONS}) is necessary. The resulting cancellation depends on the relative {\em sign} of the eigenvalues $M_{\tilde\chi^-_A}$ of $M_C$ and on the products of the matrix elements $(O_R)_{\sss A1}(O_L)_{\sss A2}$ for $A=1,2$. We report in tab.~\ref{tab:SIGNS} the signs of the various contributions entering eq. (\ref{eq:CHRGCONTRIBUTIONS}). Notice that the overall contributions have a minus sign in front.
\begin{table}
\begin{center}
\begin{tabular}{c|c|c|c|c|}
 & $M_2<-\mu$ & $-\mu<M_2<0$ & $0<M_2<\mu$ & $M_2>\mu$\\
\hline
\hline
$(O_R)_{\sss 11}(O_L)_{\sss 12}$ &$-$&+&+&+ \\
$M_{\tilde\chi^-_1}$ &+&$-$&+&+\\
\hline
{\bf A=1} &+&+&$-$&$-$\\
\hline
\hline
$(O_R)_{\sss 21}(O_L)_{\sss 22}$ &+&$-$&+&+\\ 
$M_{\tilde\chi^-_2}$ &+&$-$&$-$&$-$\\
\hline
{\bf A=2} &$-$&$-$&+&+\\
\hline
\hline
{\bf Charg.} &$-$&$-$&+&+\\
\hline
\hline
\end{tabular}
\end{center}
\caption{The sign of the terms appearing in the chargino contribution of eq.~(4.1).}
\label{tab:SIGNS}
\end{table}
For $r_2$ far from zero, the difference between the $A=1,2$ contributions, which is roughly constant even in the region close to $M_2=\pm\mu$, is driven by the combination $\sum_A(-1)^A(O_R)_{\sss A1}(OL)_{\sss A2}$. While the sign flip is due to the change of the sign of $M_2$, the peaks at $r_2\approx0$ are due to the sudden decrease of the particle mass  spectrum. 
We stress that the absence of a peak at $M_2=\pm\mu$ is entirely due to the particular arrangement of signs shown in tab.~\ref{tab:SIGNS}. An analogous cancellation takes place also in the case of the neutralino, though the more involved analytics make it less transparent to elucidate.

\begin{figure*}
\begin{center}
\begin{tabular}{ccc}
\includegraphics[scale=0.5]{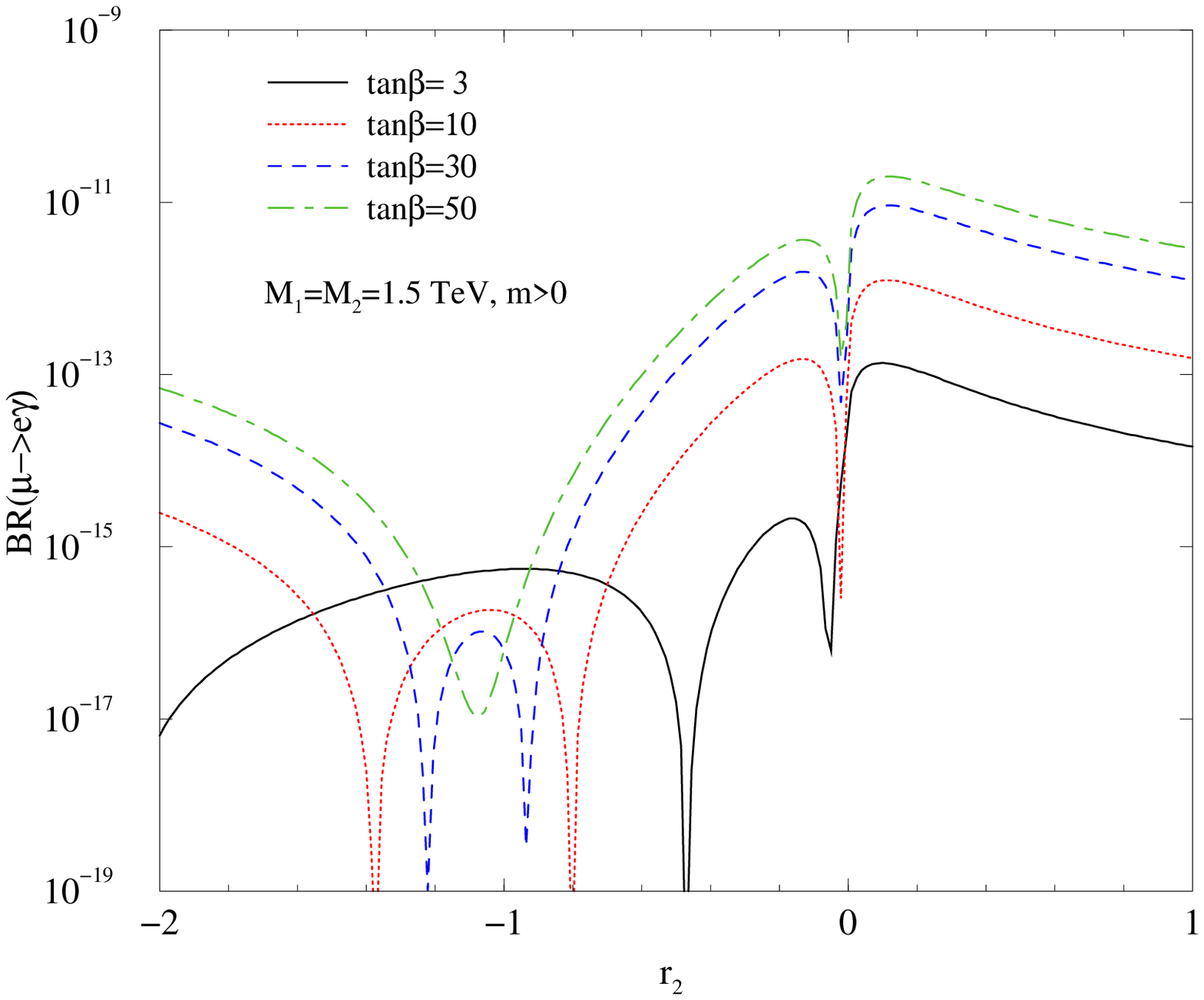}&\qquad&
\includegraphics[scale=0.5]{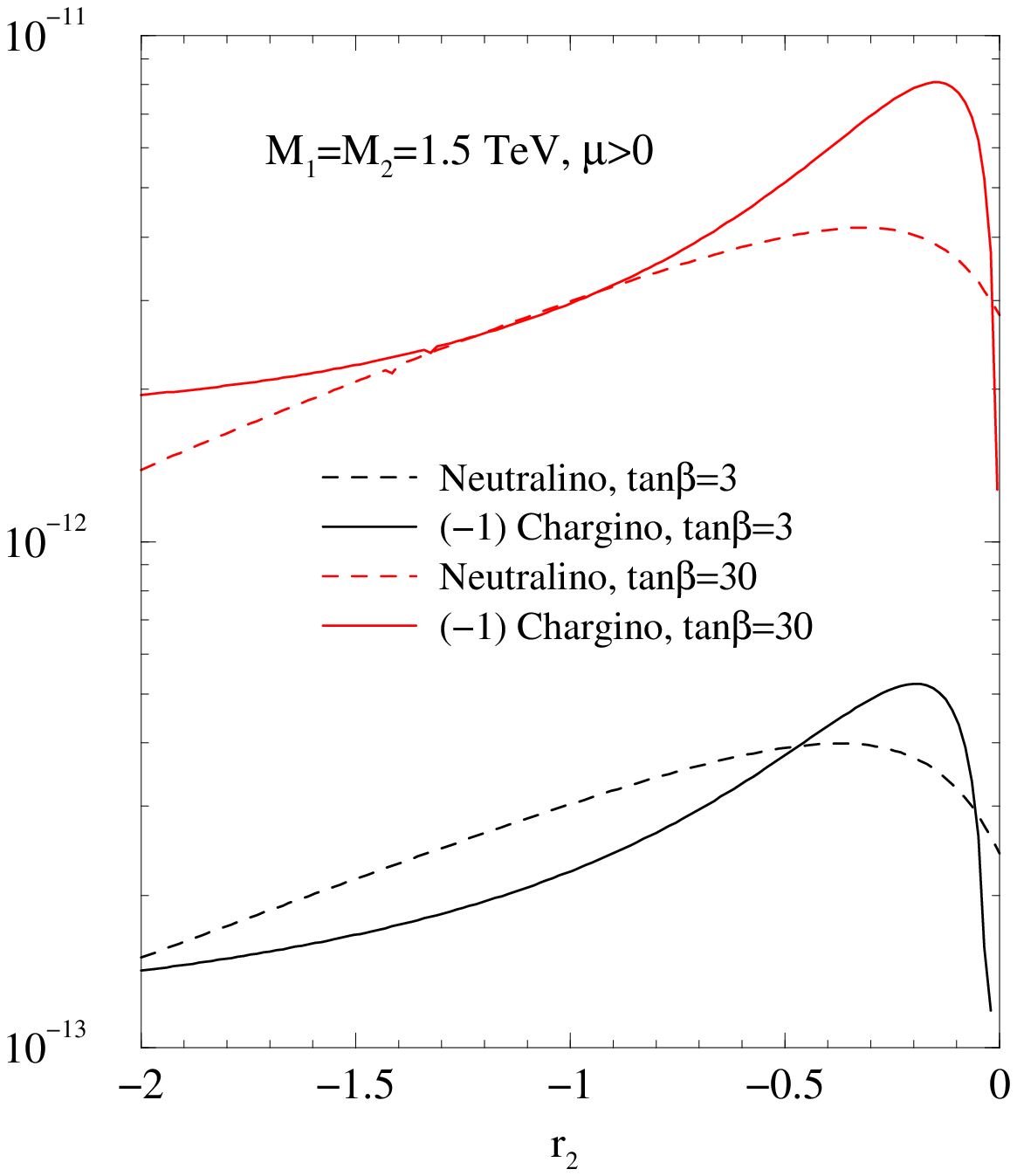}\\
\end{tabular}
\caption{\label{fig:TANB} ({\em Left} ): The $\brmu$ as a function of $r_2$ at $M_1=M_3=1.5\ {\rm TeV}$, $\mu>0$ and $m_0=1000$ for $\tan\beta=3,\ 10,\ 30$ and 50. ({\em Right} ): The details of chargino and neutralino contributions for $\tan\beta=3$ and 30. The supersymmetric parameters are the same as for the left panel.}
\end{center}
\end{figure*}

Finally, we comment on the dependence of the outlined structure for lepton flavor violation on the other relevant supersymmetric parameters. First, taking a different value for $\tan\beta$ changes the position of the dips, as highlighted in the left panel of fig.~\ref{fig:TANB}. This is due to the fact that the chargino contribution is {\em proportional} to $(1/\cos\beta)\approx\tan\beta$, while the neutralino-slepton amplitude is proportional to terms of the type $(a+b/\cos\beta)$ \cite{hisano}: as $\tan\beta$ increases, the neutralino contribution grows therefore {\em more slowly} than the chargino contribution. Though this effect is rather small, the position of the dips can substantially change, as showed in the right part of fig.~\ref{fig:TANB}. 

\begin{figure*}
\begin{center}
\includegraphics[scale=0.5]{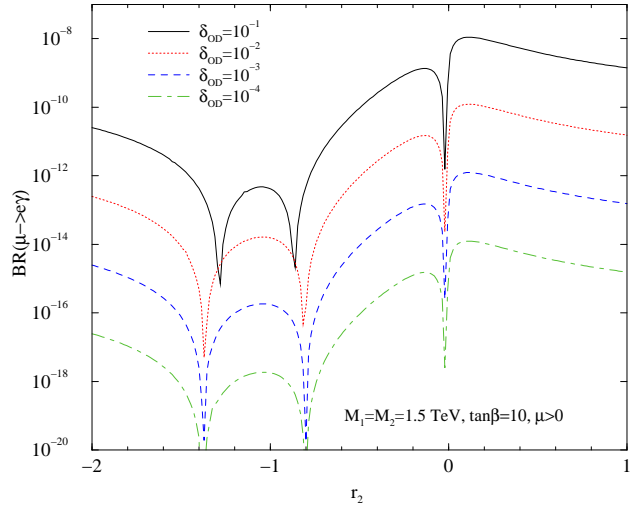}\\
\caption{\label{fig:DEL} The $\brmu$ as a function of $r_2$ at $M_1=M_3=1.5\ {\rm TeV}$, $\mu>0$, $\tan\beta=10$ and $m_0=1000$ for various $\delta_{\sss\rm OD}=10^{-1},\ 10^{-2},\ 10^{-3},\ 10^{-4}$.}
\end{center}
\end{figure*}

As far as the dependence on the relative size of the off-diagonal entries of $m_{\tilde L}$, is concerned, we study in fig.~\ref{fig:DEL} the variations induced in $\brmu$ by changing, in a range of four orders of magnitude, the size of $\delta_{\sss\rm OD}$. We find, as expected, that $\brmu\propto\delta^2_{\sss\rm OD}$ to a good accuracy. Moreover, increasing $\delta_{\sss\rm OD}$ slightly enhances the chargino contributions with respect to those of the neutralino, so that the dips tend to get closer to each other. Nevertheless, no critical dependence on $\delta_{\rm\sss OD}$ is found in the locations of the dips within a span of four orders of magnitude.  

\section{Non-universal Gluino Mass Term $M_3$}
Since gluinos do not take part in the loops of LFV decays, the gluino mass term can {\em only indirectly} affect the process $\mu\rightarrow e\gamma$. Indeed, the electroweak symmetry breaking condition gives, at $\tan\beta=10$,
\begin{equation}\label{ewsb}
\mu^2+\frac 12 m_Z^2\approx -0.1 m_0^2+2.1 M_3^2-0.22 M_2^2+0.19 M_2M_3+0.03 M_1M_3,
\end{equation}
and the coefficients vary rather mildly over the moderate $\tan\beta$ region.
We see that the dominant contribution to the right-hand side comes from $M_3$; hence, $\mu$ is practically {\em determined} by the gluino mass. Notice also that for small values of $M_3$, $\mu^2$ becomes negative, thus signalling the absence of radiative electroweak symmetry breaking.


\begin{figure*}
\begin{center}
\begin{tabular}{ccc}
\includegraphics[scale=0.5]{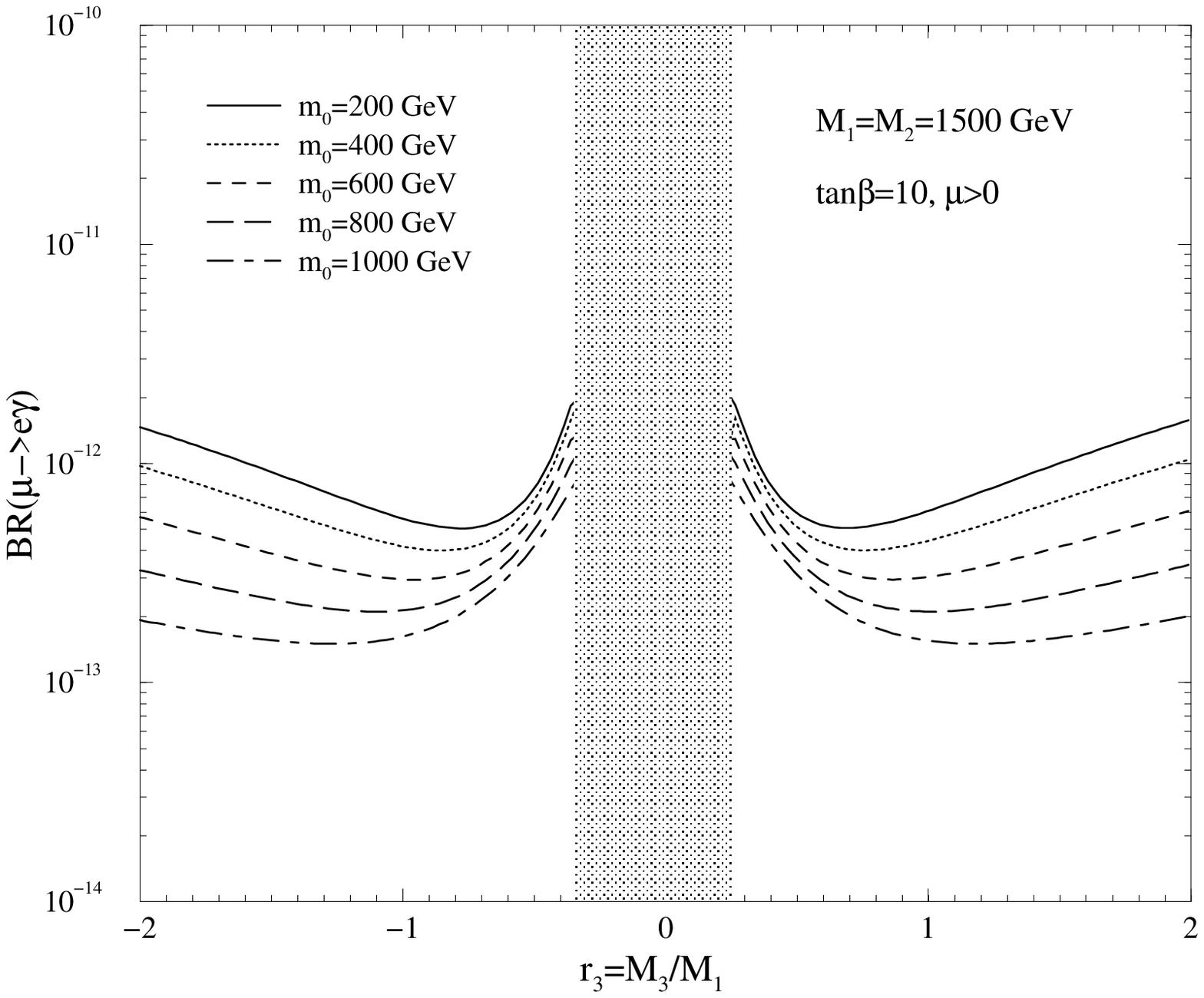}&\qquad&
\includegraphics[scale=0.5]{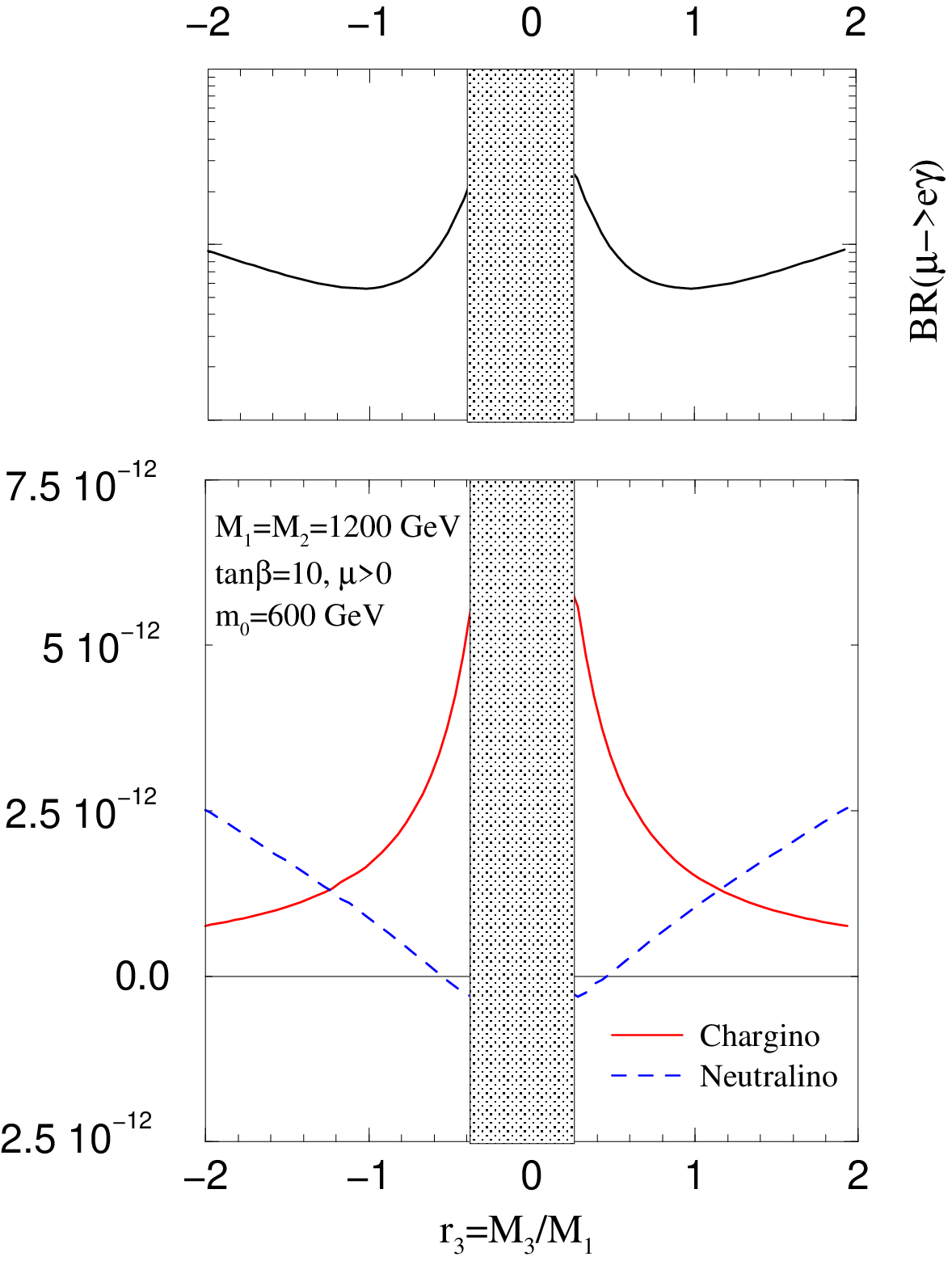}\\
\end{tabular}
\caption{({\em Left} ): The $\brmu$ as a function of $r_3$ at various $m_0$. In the figure $M_1=M_2=1500$ GeV, $\tan\beta=10$ and $\mu>0$. ({\em Right} ): The $\brmu$ as a function of the gluino high energy input value $M_3$, parametrized by the quantity $r_3=M_3/M_1$ ({\em upper panel} ) and the separate contributions from the chargino (solid line) and neutralino (dashed line) loops ({\em lower panel} ). For definiteness, we set $M_1=M_2=1200$ GeV, $\tan\beta=10$, $\mu>0$ and $m_0=600$ GeV.}\label{BRH}\label{cnr3}
\end{center}
\end{figure*}
 In the left panel of fig. \ref{BRH} we show $BR(\mu\rightarrow e\gamma)$ as a function of $r_3$ for different values of $m_0$ and $\mu>0$. In agreement with eq.(\ref{ewsb}), the branching ratio is symmetric with respect to $r_3\leftrightarrow -r_3$. Hence, we can limit our analysis to the case $r_3>0$. We notice that the behavior of the branching is not monotonous. At small values of $r_3$, $BR(\mu\rightarrow e\gamma)$ decreases until it reaches a point $r_c$ where it begins to increase. Larger values of $m_0$ imply larger values of $r_c$.

In order to understand this peculiar behavior, we have split, in the right panels of fig.~\ref{cnr3}, the chargino and the neutralino contributions. It is clear from the figure that at small $r_3$ the branching ratio is dominated by the chargino contribution, which decreases with $r_3$.  At larger values of $r_3$ the neutralino contribution dominates,  increasing with $r_3$. $r_c$ is the point where these two contributions become equal. We emphasize that such behavior is rather surprising, because at large $r_3$ the particle spectrum becomes {\em heavier}, and one would therefore expect, instead, some kind of suppression in $\brmu$.
\section{Analysis of $\brmu$ in Specific Models}
Until now, we have studied generic models with non-universal gaugino masses. In this section, we consider two well-motivated specific models which constrain soft-breaking parameters and predict  explicit relations between the three gaugino masses: $SU(5)$ GUT inspired gaugino non-universality and minimal gaugino mediation.

\subsection{SU(5)}
In supergravity models, the gauge kinetic function depends on a chiral superfield $\phi$, whose auxiliary $F$-component acquires a large vacuum expectation value. Gaugino masses come from the following five dimensional operator
\begin{equation}
L=\frac{\left<F_\phi\right>}{M_{\sss\rm PL}}\lambda_i\lambda_j,
\end{equation}
where $\lambda_{1,2,3}$ are the bino $\widetilde B$, the wino $\widetilde W$ and the gluino $\widetilde g$ fields respectively. In models based on $SU(5)$, gauginos belong to the adjoint representation of $SU(5)$, and $F_{\phi}$ can belong to any irreducible representation appearing in their symmetric product,
\begin{equation}
(24\times 24)_{\rm sym}=1+24+75+200\,.
\end{equation}
The mSUGRA model assumes $\phi$ to be a singlet, which implies equal gaugino masses at the GUT scale. On the other hand, if $\phi$ belongs to one of the non-singlet representations of $SU(5)$, gaugino masses are not universal, and they are related to one another via representation invariants \cite{ref:gaugnonuniv}. The resulting ratios for the gaugino masses at the GUT scale \cite{Anderson:1996bg} are  shown in Table \ref{table}. Some phenomenological implications of these models were analyzed, {\em e.g.}, in \cite{Huitu:1999vx} and \cite{Chattopadhyay:2003yk}. 
\begin{table}[b]
\begin{center}
\begin{tabular}{|c|ccc|}\hline
rep & $M_1$&$M_2$& $M_3$\\\hline\hline
1 & 1 &1 &1 \\
24& -1/2& -3/2& 1\\
75 & -5 & 3&1\\
200 & 10 &2 &1\\\hline
\end{tabular} 
\end{center}
\caption{Relative values of the gaugino masses at the GUT scale for the different representations allowed in $SU(5)$}
\label{table}
\end{table}

We show the behavior of $\brmu$ as a function of $M_3$ for the different representations of $SU(5)$. In this case we do not find any cancellation. For the representations $24$ and $200$ the chargino and neutralino contributions have the same sign, so that they always interfere constructively. In the representation $75$, instead, the signs are opposite, but the chargino dominates over the whole parameter space. This, however, does not necessarily imply that such cancellations do not take place in a {\em generic} $SU(5)$ GUT, because the gauge kinetic function can involve several chiral superfields belonging to different representations, and the final gaugino masses would be linear combinations of the different contributions.
\begin{figure*}
\begin{center}
\includegraphics[scale=0.5]{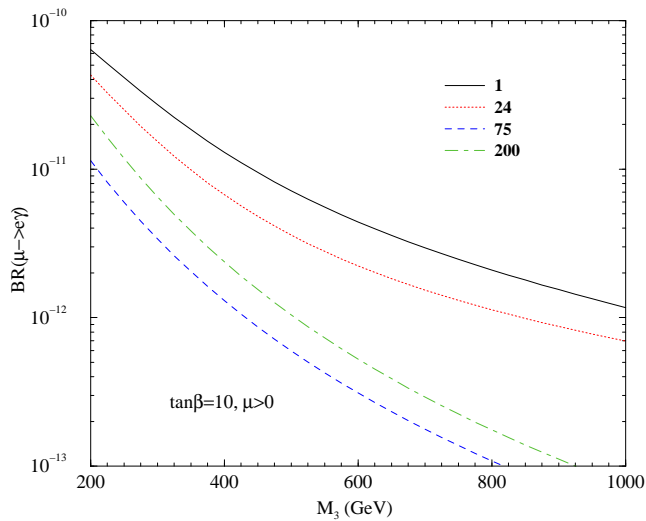}
\end{center}
\caption{Non universal gaugino masses with high energy ratios lying in the representations of the symmetric product of two $SU(5)$ adjoints. In this plot we set $m_0=1200$ GeV, $\mu>0$ and $\tan\beta=10$.}\label{su5}
\end{figure*}

\subsection{Gaugino Mediation}

Models with gaugino mediated symmetry breaking naturally emerge in higher dimensional GUT theories \cite{gauginomediation}. In such theories, the MSSM matter fields are localized on a \emph{visible} brane, while the gauge fields can propagate in the bulk of the extra dimensions. Supersymmetry is broken on a distant \emph{hidden} brane, separated from the visible one in the extra dimensions. Extradimensional locality forbids direct couplings between the two branes, and therefore suppresses all soft masses which involve MSSM matter fields (squark and slepton masses as well as trilinear couplings). Gauginos, on the contrary, can directly couple to the source of SUSY breaking, acquiring nonzero masses. As a result, below the compactification scale, $M_c\sim 1/R$, the effective 4-dimensional theory is the MSSM, and gaugino masses are the only non-negligible sources of SUSY breaking. At low energies, scalar masses and trilinear couplings are generated from gaugino masses through renormalization group running. 

From the phenomenological point of view, models with gaugino mediation are very attractive. On the one hand they  alleviate the flavor and $CP$ problems in supersymmetric theories, on the other hand, in a context of grand unification, the doublet-triplet splitting of Higgs fields can be easily achieved, and proton decay through dimension five operators can be naturally suppressed.  

\begin{figure*}[h!]
\begin{center}
\includegraphics[scale=0.5]{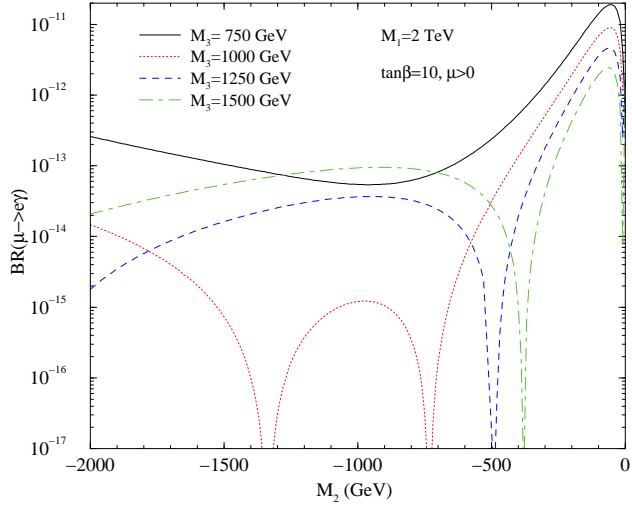}
\end{center}
\caption{$\brmu$ in a model with gaugino mediation and non-universal gaugino masses at $M_c=M_{\sss\rm GUT}$. $M_1$ is set to 2 TeV, $\tan\beta=10$ and $\mu>0$. We plot the results as functions of $M_2$ at four different values for $M_3=$ 750, 1000, 1250 and 1500 GeV.}\label{GM}
\end{figure*}
\begin{figure*}[h!]
\begin{center}
\includegraphics[scale=0.5]{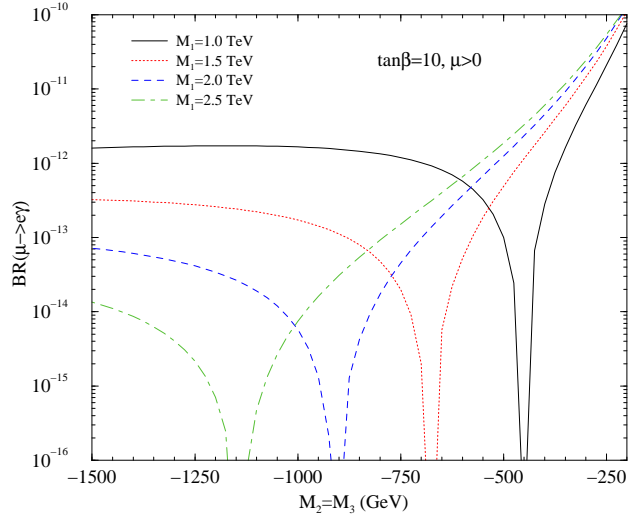}
\end{center}
\caption{$\brmu$ in models with gaugino mediation and restricted flipped $SU(5)$ gauge symmetry, which enforces non-universal gaugino masses and dictates $M_2=M_3$.}\label{flip}
\end{figure*}
In these scenarios, at the scale $M_c$, which we assume to be the unification scale ($M_c=M_{\sss\rm GUT}$), the only free parameters are the gaugino masses $M_i$ and  $\tan\beta$, since $m_0$ and $A_0$ are set to zero. In order to avoid a charged LSP (the lightest stau) one has either to resort to some specific GUT running above $M_c$ in the universal case \cite{Profumo:2003sx}, or to gaugino non-universality \cite{Komine:2000tj,Baer:2002by,Balazs:2003mm}. In what follows, we will limit our discussion to the case, relevant for the present context, of non-universal gaugino masses. In fig.\ref{GM} we show a plot of $\brmu$ as a function of $M_2$ for fixed $M_1$ and different values of $M_3$. Once again, cancellations occur when $M_2$ and $M_1$ have opposite signs. 

If SUSY is broken on a brane with restricted gauge symmetry, non universal gaugino masses are generated. For example, if the bulk symmetry is $SO(10)$, hidden branes with Pati-Salam $SU(4)\times SU(2)_L\times SU(2)_R$, Georgi-Glashow $SU(5)\times U(1)$, or flipped $SU(5)'\times U(1)'$ gauge symmetries can be obtained \cite{Dermisek:2001hp,Barr:2002fb}. If gauginos get masses on these branes, the gauge symmetry relates the gaugino masses of the MSSM at the compactification scale. In flipped $SU(5)$, for instance, one obtains $M_3=M_2\ne M_1$ \cite{Barr:2002fb,Balazs:2003mm}. In Fig.\ref{flip} we have considered this case. We show the branching as a function of $M_2=M_3$ for several values of $M_1$. In this specific model, for any value of $M_1$ there exists a value of $M_2=M_3$ for which the exact cancellation between chargino and neutralino contributions, and the resulting suppression of $\brmu$,  occurs. As emerging from the figure, this value turns out to be an increasing function of $M_1$.

\section{Conclusions}

We studied in this paper the implications of non-universal gaugino masses on LFV decays of the type $l_i\rightarrow l_j\gamma$. The analysis of the interplay between chargino-sneutrino and neutralino-slepton loop amplitudes lead us to conclude that cancellations between the two contributions may drastically suppress $\brmu$ when non-universalities are allowed in the gaugino sector at the high energy scale $M_{\sss GUT}$. 

In particular, we found that such cancellations, which are absent in the case of universality, take place when the signs of $M_2$ and $M_1$ are opposite. Moreover, we highlighted that $\brmu$ does not always decrease when the particle mass spectrum increases, but, owing to the neutralino contribution dominance, and the above mentioned destructive interferences, it may increase as well. In particular, when $M_3\gg M_2,M_1$, $\brmu$ is found to be an {\em increasing} function of $M_3$. We showed that a peculiar sign arrangement, which we analyzed in some detail, smoothes the sharp peak which takes place in the separate contributions from the two charginos at $M_2=\pm\mu$. We also concentrated on particular models where gaugino non-universality plays a fundamental r\^ole, as in gaugino mediation, and showed that even in this case where one deals with a restricted parameter space, cancellations may greatly suppress $\brmu$. Further constraints on the ratios between the three gaugino masses, stemming from grand unification, may still allow these types of suppressions, as we demonstrated in the particular case of flipped $SU(5)$, again in the context of gaugino mediation. 

We emphasize that our approach does not depend on the particular model-dependent source of lepton flavor violation in the MSSM, as long as the outlined assumptions about the dominant source of LFV are fulfilled; the presence of subdominant contributions would not change our conclusions.

\end{document}